# Thermal spike− induced sublimation of carbon nanotubes

*Sumera Javeed[1], Shoaib Ahmad[2*]*

[1]Pakistan Institute of Nuclear Science and Technology, P O Nilore, Islamabad, Pakistan

[2]National Center for Physics, QAU Campus, Shahdara Valley, Islamabad 44000, Pakistan

[*]Corresponding author: Tel/Fax: (+9251)2700669; sahmad.ncp@gmail.com

## ABSTRACT

We report and provide justification for the consistently observed four experimental facts of the mass spectrometric data of carbon cluster $C_x (x \geq 1)$ emission from the low-energy $Cs^+$-irradiated single-walled carbon nanotubes (SWCNTs). Firstly, the diatomic carbon $C_2$ is the most abundant sputtered species for $Cs^+$ in the energy range $0.2\ keV \leq E(Cs^+) \leq 2.0\ keV$. Secondly, monatomic carbon $C_1$ is emitted with the least sputtering yield. Thirdly, at low cesium energies i.e., $E(Cs^+) < 400$ eV, the emitted species are $C_2$, $C_3$ and $C_4$. Lastly, as the irradiating $Cs^+$ energy increases, the normalized yield of $C_1$ monotonically increases while $C_2$, $C_3$ and $C_4$ show gradual decrease and saturation. The experimental data for the normalized density of clusters and atoms $n_{C_x} = \left(N_{C_x} / \sum_x N_x\right)$ follows the pattern $n_{C_2} > n_{C_3} > n_{C_4} > n_{C_1}$. Sputtering of clusters is proved here to be due to thermal spikes. Binary collision cascade theory does not explain cluster sputtering. A statistical thermal model is developed to explain the experimentally observed data.

The probability of a cluster $C_x$ to be emitted is shown to be proportional to that for the creation of an x-member vacancy with formation energy $E_{xv}$ at temperature $T_s$ as $p_x = \{exp(E_{xv}/kT_S) + 1\}^{-1}$. The energies of formation of single and double vacancies $E_1$ and $E_2$ from DFT calculations and the ratio of normalized experimental yields $(n_{C_1}/n_{C_2})$ have been used to estimate $T_s$ in the ratio $p_1/p_2 \approx n_{C_1}/n_{C_2} = \{exp(E_2/kT_S) + 1\}/\{exp(E_1/kT_S) + 1\}$. For given values of the formation energies and using the experimental ratios $n_{C_1}/n_{C_2}$ we can determine the locally enhanced spike temperature $T_S$. Once $T_s$ is evaluated, the formation energies of tri- and quarto-vacancies are obtained by using the ratios of normalized densities $n_{C_2}/n_{C_3}$ and $n_{C_2}/n_{C_4}$. We show that by invoking thermal spikes, cluster emission from, and the multiple vacancy generation in, the $Cs^+$−irradiated SWCNTs can be explained. We also suggest modifications to Monte Carlo type calculations of sputtering.

## I. INTRODUCTION

We have conducted a series of experiments in order to investigate the nature of the constituents of the sputtered species emitted from the irradiated nanostructures of carbon [1-3]. The emitted species have consistently provided data of clusters that outnumber the monatomic yields. The irradiations have included intense, pulsed electrons and ions and continuous $Cs^+$ beam. The targets were fullerite, single and multi-walled carbon nanotubes. For the sake of comparison, graphite had been the control target for all the experiments. Here we choose to explain the results from cesium irradiated SWCNTs. The results and conclusions drawn from the single shelled nanotubes can be extended to the irradiated MWCNTs and graphene sheets. Cesium ion has been chosen as the irradiating projectile as its energy, intensity and dose can be accurately controlled and monitored in the source of negative ions with cesium sputtering (SNICS). The sputtered species are emitted predominantly as neutrals but delivered as negatively charged anions by the source. In SNICS one avoids the high temperatures that are associated with plasma sources for such studies. Survival of the sputtered large carbon clusters is much higher in the relative low temperatures associated with SNICS compared with those in hot plasmas. Similar sputtering experiments with $Xe^+$ irradiation of graphite had identified significantly higher emissions of negative clusters $C_x^{-1} (x \geq 1)$ [4].

Two fundamental questions arise when analyzing the mass spectra of carbon anions sputtered from $Cs^+$-irradiated single walled carbon nanotubes:

1. Why $C_2$ is the main sputtered species rather than the monatomic $C_1$?

2. Which mechanism explains the emission and dominance of multi-atomic carbon clusters?

Linear collision cascade–based sputtering theories [5-8] do not explain the sputtering of clusters. The atomic collisions cascade that evolve in the sub-surface layers under the irradiated surface, have been proposed for the sputtering of atomic constituents. Collision cascade theory predicts sputtering yields by utilizing ion to target mass ratio, angle of ion incidence, energy of the incident ion as the parameters. Energy of direct recoils is shared in the cascades of binary collisions. Sputtering yield, therefore, counts the number of recoiling target atoms that leave the outer surface per ion. Clusters are neither considered nor sputtered from the cascades. Furthermore, due to the mono-layered nature of single walled carbon nanotubes the spread of cascades is restricted to planar geometries.

A statistical thermal model is developed based on the conjecture that at lower $Cs^+$ energies ($\leq$ few keV) $Cs^+$−nanotube collisions are the most efficient means of energy transfer rather than the Cs−C atom collisions. The share of the phonons in the energy transferred to the lattice is significant. Thermal spike may develop where the local temperature $T_s$ is high enough for the sublimation of the constituents of the nanotube, atoms as well as the clusters.

We compare our experimental results and those from our thermal spike model with SRIM [9] generated data and information for similar irradiations and target geometries. Since SRIM generated collision cascades are statistical distributions of ionic collisions that do not account for steady state conditions between ion-atom collisions events. Sputtering of atoms is treated as a statistical event caused by the interaction of cascades with surface. We have tried to trace the origin of the cluster sputtering, in the energy invested in phonons. SRIM's data on the ratio of phonon energies seems relevant to our discussion.

## II.  THE EXPERIMENT

We have chosen SNICS as the source that was designed to use heavy metallic ion $Cs^+$ to irradiate targets to sputter their constituents and extract these as anions [10]. SWCNTs of 2 nm diameter, 3-13 μm length were compressed in Cu bullets used as targets for the source (SNICS) installed on 2 MV Pelletron at GCU, Lahore. An extended sequence of experiments was conducted with SWCNTs with $Cs^+$ energy varied from 0.2 to 2.0 keV with increments of 0.1 keV. The source is the key ingredient of our experiments as it is able to produce stable $Cs^+$ beam for extended periods at a given energy in the range from 200 eV to 5.0 keV. The sputtered and recoiling atoms and the emitted clusters get negatively charged while leaving the target surface and are extracted as anions. Energy of the extracted species is determined by the target and the extraction voltages. The sputtered species are extracted at the constant energy of 30 keV. A momentum analyzer was used to collect mass spectra of the sputtered species as a function of the cesium energy $E(Cs^+)$. Cluster number densities were determined from areas under respective peaks from the mass/charge (m/z) spectra as a function of $E(Cs^+)$.

Nineteen (19), consecutive sets of mass spectra for the sequence of irradiations with successively increasing $E(Cs^+)$ were recorded. The purpose of experiments was to retain the pristine nature of SWCNTs for irradiations with low energy to identify the nature of the sputtered species at minimum damage to the nanotubes. Even with low intensity $Cs^+$ beams, each spectrum is obtained in 300 s and the SWCNTs were exposed to increasing energy and intensity of $Cs^+$ beam for irradiation time ~ 6000 s. The experiments on heavily irradiated SWCNTs show variations in their cluster emissions but the essential features of the data are not significantly different [3].

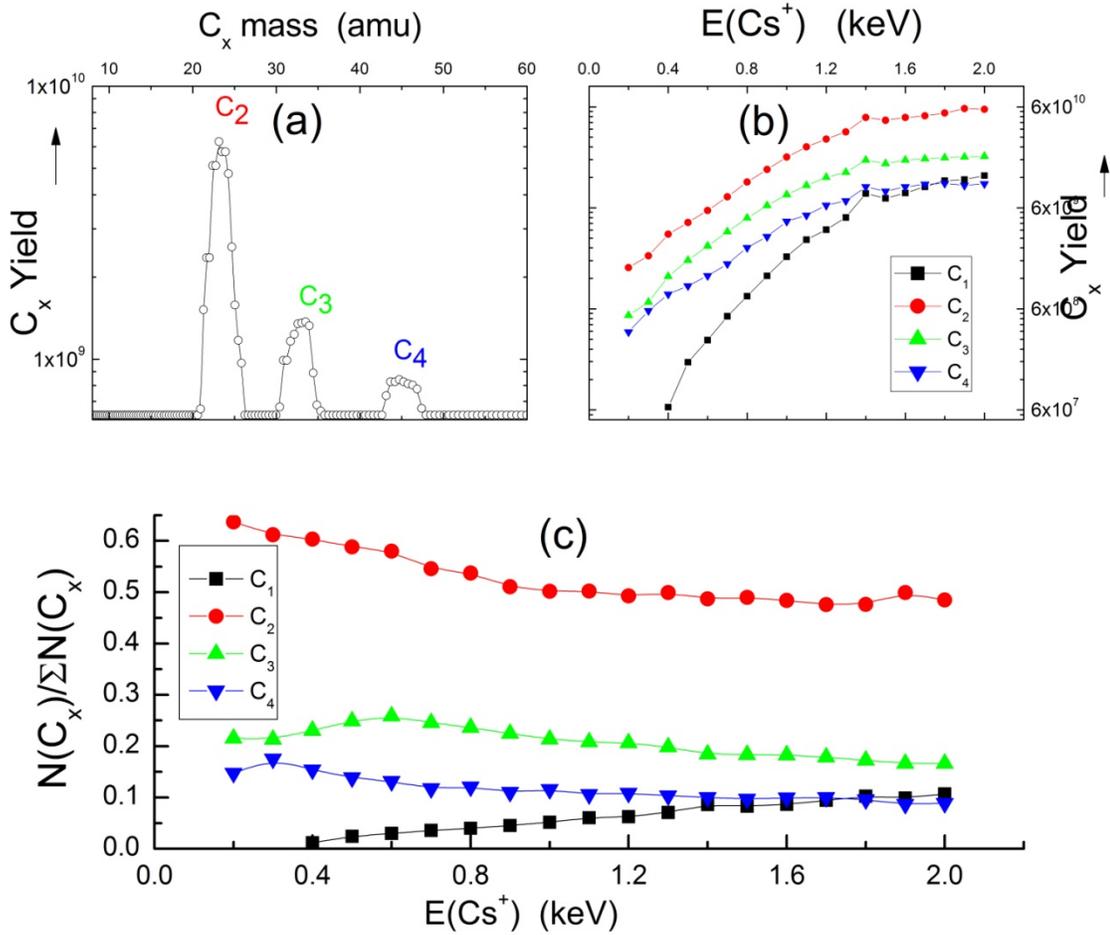

Figure 1. (a) The first mass spectrum of carbon cluster anions $C_x$ emitted from SWCNT when irradiated with $E(Cs^+) = 0.2$ keV. (b) The yield of anions $C_x$ is plotted as a function of $E(Cs^+)$ in the energy range 0.2 to 2.0 keV. The yield axis is logarithmic. (c) The normalized yield $n_x=N(C_x)/\Sigma N(C_x)$ are plotted to identify the profile of clusters and $C_1$ as a function of increasing $E(Cs^+)$.

Experimentally determined values of the normalized yields of sputtered clusters are obtained from the mass spectra of clusters emitted from the $Cs^+$-irradiated SWCNTs as a function of its energy $E(Cs^+)$. One such spectrum at $E(Cs^+) = 200$ eV is shown in Fig.1(a) which shows three peaks due to $C_2$, $C_3$ and $C_4$. This is the first spectrum from the series. It deals with the pristine, undamaged nanotubes. Results for the areas under each cluster peak at increasing $Cs^+$ energy are shown from nineteen (19) spectra in Fig. 1(b). The normalized yield for the ejection of a cluster with x-atoms $C_x$ is obtained from each mass spectrum by summing over all the cluster yields ($\Sigma N_{C_x}$) and then normalizing for each species; it provides $n_x = N_{C_x}/\Sigma N_{C_x}$. This is the experimentally measured probability of emission of a particular cluster species $C_x$. Fig. 1(c) has the nineteen (19) sets of the normalized yields or densities $n_x$ plotted as a function of $E(Cs^+)$ for the four emitted species $C_2$, $C_3$, $C_4$ and $C_1$.

## III. THE MODEL

Let us consider a monolayer of N carbon atoms as a graphene sheet or in cylindrical form as single-walled nanotube. We assume that thermal spikes are initiated in localized regions that are at elevated temperatures $T_s$. Monatomic or multi-atomic vacancies can be created if one, two, three, four or x-numbers of carbon atoms are removed in the form of clusters $C_x$. Single carbon atoms or clusters $C_x$ are bonded to the matrix of surrounding C atoms with binding energy $E_x$. When $n$ vacancies with $x$-C atoms are created in a target with N carbon atoms, the number of ways this can be done is

$$W = N!/(N-n)!\,n!$$

The associated entropy is $S_x = k \ln W$. The internal energy is $U = nE_x$ and temperature $T_s$ is related to $E_x$ and entropy $S_x$ is $1/T_s = \frac{k}{E_x}(\partial W/\partial n)$. This leads to

$$\frac{n}{N} = \{(\exp(E_x/kT_s) + 1\}^{-1} = p_x \qquad (1)$$

It is the probability of creation of a vacancy of $x$-C atoms.

Experimentally determined values of the normalized yields of the sputtered clusters $C_x(x \geq 1)$ are obtained from the mass spectra of clusters emitted from the $Cs^+$-irradiated SWCNTs as a function of Cs energy. The normalized yields are $n_x = N_{C_x}/\Sigma N_{C_x}$. Since $n_x$ is the experimentally measured density of emission of $C_x$, therefore it is directly proportional to the probability of thermally created vacancies $p_x$. The probabilities of emission of any of the cluster species, for example $C_2$, $C_3$, $C_4$ and higher ones, is $p_x = n_x/N_S$, where $N_S$ is the total number of C atoms in the spike region. Since $N_S$ is an unknown quantity, therefore, we utilize the ratio of the probabilities that eliminates $N_S$. The ratio of probabilities allows calculation of physical quantities and variables like energies of formation of respective vacancies $E_{xv}$ and the spike temperature $T_S$. These ratios are

$$p_x/p_y = \frac{n_x}{n_y} = \{exp(E_2/kT_S) + 1\}/\{exp(E_1/kT_S) + 1\} \qquad (2)$$

$C_1$ will be included in application of thermal spike model along with multi-atomic clusters. But as we will see its density profile as a function of $E(Cs^+)$ is different from those of the clusters ($C_2$, $C_3$ and $C_4$). This discrepancy is discussed and explained below.

## A. Thermal spikes and collision cascades

The model developed in this communication is validated by the experimental data presented in Fig.1. We compare its findings with the typical sputtering experiment in the same energy range using the Monte Carlo statistical model of SRIM [9]. For this purpose, we have tried to simulate our experiment shown in the inset of Fig. 2 in a comparable SRIM model, also shown in the inset. The figure for the experiment shows a 2 nm diameter nanotube being irradiated with $Cs^+$ ions. The individual C atoms and the nanotube as a whole, offer a whole range of collision cross sections; from head on to glancing collisions. These are simulated by assuming equivalence of ion−graphene geometries shown under SRIM in the inset. The $Cs^+$ incident angle varies from 0° to 89°. The SRIM variable α is to simulate conditions of perpendicular incidence (α=0°) on SWCNT and up to those of grazing angle collisions (α=89°). This allows us to get SRIM calculated sputtering yield of $C_1$ and the ion-recoil energy spent as phonon energy.

In Fig. 2(a) the cumulative sum of the number densities of all emitted species $C_x^{-1}(x \geq 1)$ as $\Sigma n(C_x)$ is plotted as a function of $E(Cs^+)$. It shows an early quadratic increase up to $E(Cs^+) \leq 1$ keV followed by a linearly increasing pattern. The SRIM result for sputtering yield of C atoms $S(C_1)$ is shown in Fig. 2(d). Comparing the two data for $\Sigma n(C_x)$ and $S(C_1)$ in Fig. 2(a) and 2(d) one can see that the SRIM simulation is (a) missing the mechanism for cluster emission and (b) assuming that the only way the energy can be released from the $Cs^+-$ irradiated monolayer is by emitting $C_1$.

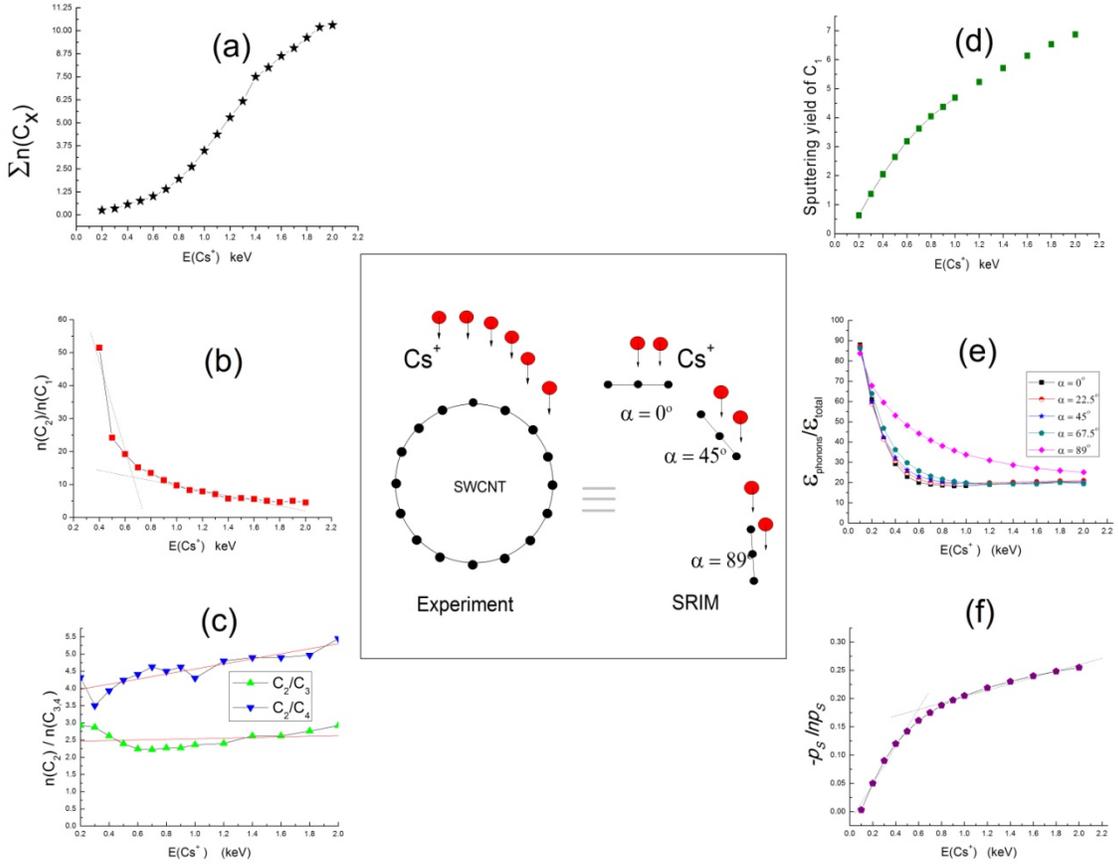

Figure 2. The inset shows the Cs−irradiated SWCNT offering broad range of collision cross sections; these are simulated by equivalent ion−graphene geometries shown in the inset. The $Cs^+$ incident angle varies from 0° to 89°. (a) The cumulative sum of the number densities of all emitted species $C_x^{-1}(x \geq 1)$ as $\Sigma n(C_x)$ is plotted as a function of $E(Cs^+)$. (b) $C_2$ being the most abundant and stable cluster in the mass spectra of the sputtered species; the ratio of the normalized yields of $C_2$ and $C_1$ is shown as a function of $E(Cs^+)$. (c) The ratios of normalized yields of $C_2/C_3$ and $C_2/C_4$ are plotted for a comparison of the three ratios. (d) The SRIM result for sputtering yield of C atoms $S(C_1)$ is shown. (e) The data from five angles of increasing $Cs^+$ incidence shows the percentage of total energy into phonons. (f) Sputtering entropy $-p_S \ln p_S$ as a function of $E(Cs^+)$ is plotted to identify the missing information of sputtering.

In Fig. 2(b) and 2(c) the ratios of normalized yields of $C_2/C_1$, $C_2/C_3$ and $C_2/C_4$ are plotted against $E(Cs^+)$. These ratios clearly indicate the emission profiles of the four emitted species. In Figure 2(b) the ratio $C_2/C_1$ varies from just over 50 at 0.4 keV to under 10 at 0.7 keV. Thereafter, the ratio shows 20% variation. The ratios of number densities of $C_2$ to $C_3$ shown in Figure 2(c) remain stable around 2.5±0.25 and 4.5±0.5 for $C_2/C_4$, for the entire range of $Cs^+$ energies i.e., from 0.2 to 2.0 keV. Whereas, $C_1$ can be emitted from an irradiated SWCNT from the binary collision cascades, clusters like $C_2$, $C_3$ and $C_4$ may only originate in a spike region.

Although Monte Carlo type simulations do not predict sputtering of clusters exclusively, these can be used to infer the higher percentages of recoil energy into phonons. Fig. 2(e) shows the percentage of total energy into phonons. The data from five angles of increasing $Cs^+$ incidence are shown. Two well defined regimes are visible, one dominated by energy dissipation into phonons $E(Cs^+) \leq 0.6$ keV and the other where binary collision sequences seem to dominate ($E(Cs^+) \geq 0.6$ keV).

From SRIM simulation presented in Fig. 2(d) we derive the probability of a $C_1$ being sputtered at ion energy $E(Cs^+)$, irradiating at angle $\alpha$ as shown in the inset. This probability is $p_S = \left(\sum_{\alpha=0°}^{89°} S_\alpha\right) / \left(\sum_\alpha \sum_E S_{\alpha,E}\right)$ where $\alpha$ is the beam-target angle and E is the irradiating energy $E(Cs^+)$.

Fig. 2(f) plots the sputtering entropy in the form of $-p_S \ln p_S$ as a function of $E(Cs^+)$. For each irradiation energy the probability was determined. It is a measure of the ignorance of the physical process—sputtering of $C_1$. It clearly shows two regions, one for $E(Cs^+) \leq 0.6$ keV and the other with $E(Cs^+) \geq 0.6$ keV. In the low energy regime, the sputtering probability is low, gradually increasing. The process has similarity with the enhanced energy dissipation into phonons in this energy regime. In Fig. 2(f) we have the region of missing information as the region of highest cluster emissions.

If one was to plot a normalized sputtering yield $S' = (S_E)/(\sum S_E)$ then it shows a clear hump around $E(Cs^+) = 0.6$ keV; $S'$ increasing to a maximum value at 0.6 keV and then decreasing at $\geq 0.6$ keV.

## B. Thermal spike temperature $T_S$

Since we only know the energies of formation of single and double vacancies, these can help us to calculate $T_S$. Energies of formation of the tri- and quarto-vacancies can later be calculated by using the $T_S$ values. The normalized yields of the sputtered clusters are obtained from the mass spectra of clusters emitted as $n_x = N_{C_x}/\Sigma N_{C_x}$, that are plotted in Fig. 2(c) for the four dominant species $C_2$, $C_3$, $C_4$ and $C_1$. These are proportional to the respective probabilities of generation of vacancies px.

We use values of the energies of formation of single and double vacancies $E_{SV}$ and $E_{DV}$ obtained from DFT and DFT-based TB calculations for single-walled nanotubes [11-15]. These are $E_{SV}$=6.8±1.0 eV and $E_{DV}$=4.7±0.5 eV. In the first stage of the model, the theoretically determined values of $E_{SV}$ and $E_{DV}$ are used. Spike temperature $T_S$ is obtained as a function of $E(Cs^+)$ utilizing the proportionality $p_x \propto n_x$ in eq. (2) and the relation $p_x/p_y = n_x/n_y$. As $\exp(E_x/kT_s) \gg 1$

$$T_S \cong [(E_{xv} - E_{yv})/k)][\ln(p_y/p_x)]^{-1} \qquad (3)$$

The calculated values of $T_S(p_2/p_1)$ for $C_2/C_1$ ratios are shown as open squares in Figure 3 as a function of $E(Cs^+)$. The first value is at $E(Cs^+)$=0.4 keV when $C_1$ is first observed in the mass spectra, besides $C_2$,$C_3$ and $C_4$. Values of $T_S$ based on the ratio on $C_2/C_1$ vary between 3000 K at $E(Cs^+)$=0.4 keV to 5500 K at $E(Cs^+) = 1.0$ keV. Even higher values of $T_S \sim 9000$ K are obtained for increasing $E(Cs^+)$. This shows an energy dependent $T_S$ with large variations and much higher values obtained than the boiling point of graphite. The anomalously high temperatures obtained are due to the use of a varying ratio of two sputtered species (e.g. $C_2$ and $C_1$) that may have different origins. In the second stage we consider the normalized ratios of higher clusters and their sputtering pattern as a function of $Cs^+$ energy.

The average values of tri- and quarto-vacancies obtained at the temperatures obtained from $C_2/C_1$ are $E_{TV}$=5.13±0.5 eV and $E_{QV}$=5.73±0.5 eV for $E(Cs^+)$= 0.4 to 1.0 keV. From the formation energies of $C_2$, $C_3$, $C_4$ and their experimental probabilities $p_2$, $p_3$, $p_4$, two sets of temperatures are calculated. These are shown in Figure 3 as the filled red triangles and blue squares. A starred, black curve of the average of the two temperatures is shown $<T_S>$=4016K. This is the average spike temperature for $Cs^+$−irradiated SWNTs in the energy range from 0.2 to 2.0 keV.

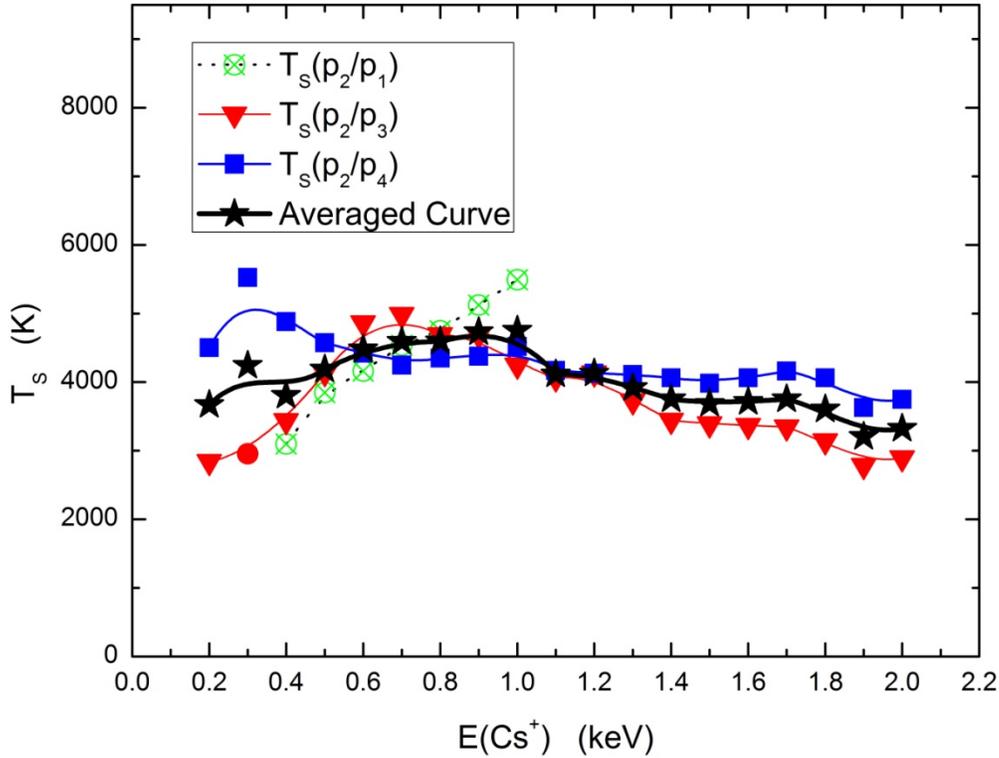

Figure 3. The spike temperature $T_S$ calculated from the normalized densities of $C_2/C_1$, $C_2/C_3$ and $C_2/C_4$ is plotted against $E(Cs^+)$. Average $<T_S>$ of $T_S(C_2/C_3)$ and $T_S(C_2/C_4)$ is shown as the thick black –starred line.

## IV. DISCUSSION

As early as 1926 von Hippel [16] had proposed 'hot spot' model of evaporation from locally heated regions of irradiated targets. The collision cascade-based theories of sputtering [5-8] eventually became well established as these could explain majority of experimental data. Sputtering yield is proportional to the energy deposited in linear collision cascades. Thermal spikes were introduced to explain the thermal origin of a fraction of sputtered particles with small energies [17-21]. Experimental evidence grew over the years [22-24]. The non-linear effect has been demonstrated in molecular and cluster bombardments.

Almost all of the experiments, theoretical and computational modeling of non-linear sputtering mentioned in the references here, dealt with 3-dimensional materials whose dimensions are much larger than the range of the bombarding ion. None dealt with mono-layered, single sheets of any material. Nanotubes of carbon offered us such a target material. Diameters of few nanometers of our SWCNTs offer restricted number of direct recoils in each nanotube thus making sputtering by collision cascades less efficient. On the other hand, this structural feature of our targets favors ion-nanotube interactions as more probable, especially at low irradiation energies which makes higher percentages of the deposited energy in phonons that raises the local temperature. The result is the lower contribution of the monatomic species and higher of the clusters. That is what we have observed and reported in this communication.

## V. CONCLUSIONS

The unique experimental results of clusters emissions from $Cs^+$-irradiated SWCNTs are explained by invoking thermal spikes. The normalized number densities of clusters ($C_2$, $C_3$ and $C_4$) and their mutual ratios show constancy against the variations of the energy of $Cs^+$ as shown in Fig. 1(c). While $C_1$, in the same figure shows an energy ($E(Cs^+)$) dependent behavior. Sputtering of clusters implied a thermal origin, and we have modeled it. The probability of the emission of a cluster containing *x*-C atoms $C_x (x > 1)$, is shown to be proportional to that of creating an *x*-missing member vacancy in the irradiated SWCNT as $p_x = \{exp(E_{xv}/kT_S) + 1\}^{-1}$. The spike temperature and the energies of formation of vacancies are evaluated by using the equivalence and proportionality of experimental data for the normalized density of clusters and atoms $n_{C_x} = \left(N_{C_x}/\sum_x N_x\right)$ and $p_x$. The calculated spike temperature is an indicator of the validity of the basic assumptions. From the experimentally determined ratio of any two of the species emitted from the irradiated SWCNT and the knowledge of their vacancy formation energies, one can calculate the spike temperature. It must remain constant for the entire range of the irradiation energies i.e. E(Cs$^+$). If it does not, then the origin of either of the two species is not thermal. This fact-checking was performed in section III-B.